\documentclass{JAC2003}

\usepackage{graphicx}
\usepackage{booktabs}
\usepackage{multicol}
\usepackage{amsmath}

\begin{document}
\title{RF field-attenuation formulae for the multilayer coating model}

\author{Takayuki Kubo\thanks{kubotaka@post.kek.jp}, Takayuki Saeki, 
High Energy Accelerator Research Organization, KEK, \\ 1-1 Oho, Tsukuba, Ibaraki 305-0801 Japan\\
Yoshihisa Iwashita, Institute for Chemical Research, Kyoto University, Uji, Kyoto 611-0011, Japan}

\maketitle

\begin{abstract}
Formulae that describe the RF electromagnetic field attenuation 
for the multilayer coating model with a single superconductor layer and a single insulator layer deposited on a bulk superconductor
are derived from a rigorous calculation with the Maxwell equations and the London equation. 
\end{abstract}

\section{Introduction}\label{section:introduction}

An idea to enhance the rf breakdown field of superconducting cavities by multilayered nanoscale coating is proposed by A. Gurevich in 2006~\cite{gurevich}.  
The model consists of alternating layers of superconductor layers ($\mathcal{S}$) and insulator layers ($\mathcal{I}$) deposited on bulk Nb. 
The $\mathcal{S}$ layers are assumed to withstand higher field than the bulk Nb and shield the bulk Nb from the applied rf surface field $B_0$, 
by which $B_0$ is decreased down to $B_i < B_0$ on the surface of the bulk Nb.  
Then the cavity with the multilayered structure is thought to withstand a higher field than the Nb cavity,   
if $B_0$ is smaller than the vortex penetration-field~\cite{beanlivingston} of the top $\mathcal{S}$ layer, and 
$B_i$ is smaller than $200\,{\rm mT}$, which is thought to be the maximum field for the bulk Nb. 
In order to evaluate the shielded magnetic field $B_i$, 
the magnetic field-attenuation formulae for the multilayered structure are necessary.

When a magnetic field is applied to a superconductor, the Meissner screening current is induced, which restricts the penetration of the field to a surface layer. 
This effect is often explained by applying the London equation $d^2B/dx^2=B/\lambda^2$ to a semi-infinite bulk superconductor in the region $x \ge 0$ 
with boundary conditions $B(0)=B_0$ and $B(\infty)=0$, 
where $\lambda$ is the London penetration depth and $B_0$ is the applied surface field parallel to the superconductor surface. 
The solution is written as $B(x) = B_0 e^{-x/\lambda}$, which means the penetration of the field is restricted to depth $\lambda$. 
This solution, however, is just a solution of the London equation for a special case: a solution for semi-infinite bulk superconductor with boundary conditions given above. 
For different configurations such as a multilayer coating model, 
the London equation should be solved with appropriate boundary conditions, and
solutions are generally different from $B(x) = B_0 e^{-x/\lambda}$. 
In this paper the RF electromagnetic field-attenuation formulae for the multilayer coating model are derived from 
a rigorous calculation with the Maxwell equations and the London equation.

\section{Magnetic field attenuation in the multilayer coating model}

\subsection{Model}

For simplicity, let us consider a model with a single $\mathcal{S}$ layer and a single $\mathcal{I}$ layer deposited on a bulk superconductor.  
The region $x<0$ is vacuum, 
the region I ($0 \le x \le d_{\mathcal{S}}$) is $\mathcal{S}$ layer with the London penetration depth $\lambda_1$, 
the region II ($d_{\mathcal{S}} < x < d_{\mathcal{S}} + d_{\mathcal{I}}$) is $\mathcal{I}$ layer with permittivity $\epsilon_r \epsilon_0$, 
in which $\epsilon_r$ is a relative permittivity and $d_{\mathcal{I}}$ is assumed to be zero or larger than a few ${\rm nm}$ to suppress the Josephson coupling~\cite{gurevichreview}, 
and the region III ($x \ge d_{\mathcal{S}} +d_{\mathcal{I}}$) is a bulk superconductor with the London penetration depth $\lambda_2$, 
where all layers are parallel to the $y$-$z$ plane and then perpendicular to the $x$-axis~\footnote{The layers of our model are not necessarily thin  
hence the discussion below can be applied to layers with arbitrary thickness.}. 
The applied electric and magnetic field are assumed to be parallel to the layers. 
In order to derive the RF electromagnetic field-attenuation formulae for this model,  
the Maxwell and London equations should be solved in the region I and III, 
and the Maxwell equations in the region II. 
Boundary conditions are given as continuity conditions of the electric and magnetic field at $x=d_{\mathcal{S}}$ and $x=d_{\mathcal{S}}+d_{\mathcal{I}}$.

\subsection{Electromagnetic field in the model}

In the region I, the Maxwell and London equations should be solved. 
The equations~\cite{hassan} are given by
\begin{eqnarray}\label{eq:MaxwellLondon1}
\triangle {\bf E} = \kappa^2 {\bf E} \,, \hspace{1cm} 
\triangle {\bf B} = \kappa^2 {\bf B}\, , 
\end{eqnarray}
where 
\begin{eqnarray}\label{eq:k2}
\kappa^2 = \frac{1}{\lambda^2} \Bigl( 1 -i\mu_0 \sigma_{\rm n} \omega \lambda^2 -\frac{\omega^2}{c^2}\lambda^2 \Bigr) \,,
\end{eqnarray}
$c$ is the speed of light, $\omega$ is the angular frequency, and $\sigma_{\rm n}$ is a conductivity for the normal conducting carriers. 
Assuming the typical value of $\lambda \simeq 10^{-7}\,{\rm m}$ with $\omega \simeq 10^9\,{\rm Hz}$ for rf applications, 
the third term of the right-hand side of Eq.~(\ref{eq:k2}) becomes $\simeq 10^{-12}$. 
Furthermore, assuming ${\rm Nb}$, ${\rm NbN}$, ${\rm MgB}_2$ or ${\rm Nb}_3{\rm Sn}$ as material of the superconductor, 
normal conductivities $\sigma_{\rm n}$ are less than $10^7\, {\rm S/m}$, and then the second term becomes less than $10^{-3}$. 
Thus contributions from the second and the third terms to the electromagnetic field distribution are negligible.  
Then Eq.~(\ref{eq:MaxwellLondon1}) are reduced to 
\begin{eqnarray}\label{eq:MaxwellLondon2}
\triangle {\bf E} = \frac{1}{\lambda^2} {\bf E} \, , \hspace{1cm} 
\triangle {\bf B} = \frac{1}{\lambda^2} {\bf B} \, .
\end{eqnarray}
The solutions that are consistent with Maxwell equations are given by 
\begin{eqnarray}
E_{\rm I} &=& P_{\rm I} e^{\frac{x}{\lambda_1}} + Q_{\rm I} e^{-\frac{x}{\lambda_1}}                               \, ,  \label{eq:EI} \\
B_{\rm I} &=& \frac{1}{i ck \lambda_1} (P_{\rm I} e^{\frac{x}{\lambda_1}} - Q_{\rm I} e^{-\frac{x}{\lambda_1}} )   \, ,  \label{eq:BI} 
\end{eqnarray}
where $E_{\rm I}$ and $B_{\rm I}$ are electric and magnetic field in the region I, respectively, 
$P_{\rm I}$ and $Q_{\rm I}$ are integration constants, and $k=\omega/c$. 
In the region II, the Maxwell equations should be solved, where contributions from dielectric losses to the electromagnetic field distribution are neglected. 
Then the solutions are given by
\begin{eqnarray}
E_{\rm II} \!\!\!\! &=&\!\!\!\! P_{\rm II} e^{i\sqrt{\epsilon_r} k(x-d_{\mathcal{S}})} + Q_{\rm II} e^{-i\sqrt{\epsilon_r}k(x-d_{\mathcal{S}})}                              \, , \label{eq:EII} \\
B_{\rm II} \!\!\!\! &=&\!\!\!\! \frac{\sqrt{\epsilon_r}}{c} (P_{\rm II} e^{i\sqrt{\epsilon_r}k(x-d_{\mathcal{S}})} -Q_{\rm II} e^{-i\sqrt{\epsilon_r}k(x-d_{\mathcal{S}})} ) \, , \label{eq:BII} 
\end{eqnarray}
where $E_{\rm II}$ and $B_{\rm II}$ are electric and magnetic field in the region II, respectively, and
$P_{\rm II}$ and $Q_{\rm II}$ are integration constants. 
In the region III, Eq.~(\ref{eq:MaxwellLondon2}) should be solved as same as the region I, and the solutions that are consistent with the Maxwell equations are given by 
\begin{eqnarray}
E_{\rm III} &=& Q_{\rm III} e^{-\frac{x-d_{\mathcal{S}}-d_{\mathcal{I}}}{\lambda_2}}                          \, , \label{eq:EIII} \\
B_{\rm III} &=& -\frac{1}{ick \lambda_2} Q_{\rm III} e^{-\frac{x-d_{\mathcal{S}}-d_{\mathcal{I}}}{\lambda_2}} \, , \label{eq:BIII} 
\end{eqnarray}
where $E_{\rm III}$ and $B_{\rm III}$ are electric and magnetic field in the region III, respectively, and
$Q_{\rm III}$ is an integration constant.  
The constants $P_{\rm I}$, $Q_{\rm I}$, $P_{\rm II}$, $Q_{\rm II}$ and $Q_{\rm III}$ are determined from the boundary conditions: 
continuity conditions of the electric and magnetic field at $x=d_{\mathcal{S}}$ and $x=d_{\mathcal{S}}+d_{\mathcal{I}}$. 
Then, we obtain 
\begin{eqnarray}
P_{\rm II} &=& \frac{Q_{\rm III}}{2} \Bigl( 1-\frac{1}{i\sqrt{\epsilon_r}k\lambda_2} \Bigr) e^{-i\sqrt{\epsilon_r}k d_{\mathcal{I}}}  \, ,   \label{eq:PII} \\
Q_{\rm II} &=& \frac{Q_{\rm III}}{2} \Bigl( 1+\frac{1}{i\sqrt{\epsilon_r}k\lambda_2} \Bigr) e^{+i\sqrt{\epsilon_r}k d_{\mathcal{I}}}  \, ,   \label{eq:QII}
\end{eqnarray}
from the continuity conditions at $x=d_{\mathcal{S}}+d_{\mathcal{I}}$, and 
\begin{eqnarray}
P_{\rm I} = \frac{1+i\sqrt{\epsilon_r}k\lambda_1}{2} P_{\rm II} e^{-\frac{d_{\mathcal{S}}}{\lambda_1}} 
    + \frac{1-i\sqrt{\epsilon_r}k\lambda_1}{2} Q_{\rm II} e^{-\frac{d_{\mathcal{S}}}{\lambda_1}}  \, ,    \label{eq:PI} \\
Q_{\rm I} = \frac{1-i\sqrt{\epsilon_r}k\lambda_1}{2} P_{\rm II} e^{+\frac{d_{\mathcal{S}}}{\lambda_1}} 
    + \frac{1+i\sqrt{\epsilon_r}k\lambda_1}{2} Q_{\rm II} e^{+\frac{d_{\mathcal{S}}}{\lambda_1}}  \, , \label{eq:QI}
\end{eqnarray}
from the continuity conditions at $x=d_{\mathcal{S}}$. 
Substituting Eq.(\ref{eq:PII}) and (\ref{eq:QII}) into Eq.(\ref{eq:PI}) and (\ref{eq:QI}), we obtain
\begin{eqnarray}
P_{\rm I} \!\!\! &=&\!\!\! \frac{Q_{\rm III}}{2} e^{-\frac{d_{\mathcal{S}}}{\lambda_1}} 
    \Bigl[ \Bigl( 1-\frac{\lambda_1}{\lambda_2}\Bigr) \cos\sqrt{\epsilon_r}kd_{\mathcal{I}} \nonumber \\
   &&     +\Bigl( \sqrt{\epsilon_r}k\lambda_1 + \frac{1}{\sqrt{\epsilon_r}k\lambda_2} \Bigr) \sin\sqrt{\epsilon_r}kd_{\mathcal{I}} \Big]  \, , \label{eq:PPI} \\
Q_{\rm I} \!\!\! &=&\!\!\! \frac{Q_{\rm III}}{2} e^{+\frac{d_{\mathcal{S}}}{\lambda_1}} 
    \Bigl[ \Bigl( 1+\frac{\lambda_1}{\lambda_2}\Bigr) \cos\sqrt{\epsilon_r}kd_{\mathcal{I}} \nonumber\\
   &&     +\Bigl( -\sqrt{\epsilon_r}k\lambda_1 + \frac{1}{\sqrt{\epsilon_r}k\lambda_2} \Bigr) \sin\sqrt{\epsilon_r}kd_{\mathcal{I}} \Big]  \, . \label{eq:QQI}
\end{eqnarray}
Now the integration constants $P_{\rm I}$, $Q_{\rm I}$, $P_{\rm II}$ and $Q_{\rm II}$ are expressed in terms of $Q_{\rm III}$. 
The constant $Q_{\rm III}$ can be expressed by the surface magnetic field $B_0$, which is given by 
\begin{eqnarray}\label{eq:B0}
B_0 
&=& B_{\rm I}|_{x=0} = \frac{1}{ick\lambda_1} (P_{\rm I}-Q_{\rm I}) \nonumber \\
&=&  
\biggl[ 
\Bigl( \frac{\lambda_1}{\lambda_2}\cos\sqrt{\epsilon_r}kd_{\mathcal{I}} \!-\! \sqrt{\epsilon_r}k\lambda_1 \sin\sqrt{\epsilon_r}kd_{\mathcal{I}} \Bigr) \cosh\frac{d_{\mathcal{S}}}{\lambda_1} \nonumber \\
&+&\!\!\!\!\! \Bigl( \cos\sqrt{\epsilon_r}kd_{\mathcal{I}} \!+\! \frac{\sin\sqrt{\epsilon_r}kd_{\mathcal{I}}}{\sqrt{\epsilon_r}k\lambda_2} \Bigr) \sinh\frac{d_{\mathcal{S}}}{\lambda_1}
\biggr] \!\! \times \!\!\frac{-Q_{\rm III}}{ick\lambda_1}\,.  
\end{eqnarray}
Then substituting Eq.(\ref{eq:PPI}), (\ref{eq:QQI}), (\ref{eq:PII}), (\ref{eq:QII}) and (\ref{eq:B0}) into Eq.(\ref{eq:BI}), (\ref{eq:BII}) and (\ref{eq:BIII}), 
the magnetic fields in the region I, II and III are given by
\begin{eqnarray}
B_{\rm I} \!\!\!\! &=&\!\!\!\! 
\biggl[ 
\Bigl( \frac{\lambda_1}{\lambda_2}\cos\sqrt{\epsilon_r}kd_{\mathcal{I}} - \sqrt{\epsilon_r}k\lambda_1 \sin\sqrt{\epsilon_r}kd_{\mathcal{I}} \Bigr) \cosh\frac{d_{\mathcal{S}}-x}{\lambda_1} \nonumber \\
&+& \!\!\!\!\! \Bigl( \cos\sqrt{\epsilon_r}kd_{\mathcal{I}} + \frac{\sin\sqrt{\epsilon_r}kd_{\mathcal{I}}}{\sqrt{\epsilon_r}k\lambda_2} \Bigr) \sinh\frac{d_{\mathcal{S}}-x}{\lambda_1}
\biggr] \times \frac{B_0}{D} \, , \label{eq:exBI} \\
B_{\rm II} \!\!\!\! &=&\!\!\!\! 
\biggl[ 
\frac{\lambda_1}{\lambda_2} \cos\sqrt{\epsilon_r}k(d_{\mathcal{S}}+d_{\mathcal{I}}-x) \nonumber \\
&&- \sqrt{\epsilon_r}k\lambda_1\sin\sqrt{\epsilon_r}k(d_{\mathcal{S}}+d_{\mathcal{I}}-x) 
\biggr]  \times \frac{B_0}{D} \, , \label{eq:exBII} \\
B_{\rm III} \!\!\!\! &=&\!\!\!\! 
\frac{B_0}{D} \frac{\lambda_1}{\lambda_2} e^{-\frac{x-d_{\mathcal{S}}-d_{\mathcal{I}}}{\lambda_2}} \, .  \label{eq:exBIII}
\end{eqnarray}
where the denominator $D$ is given by 
\begin{eqnarray}
D 
&=& \Bigl( \frac{\lambda_1}{\lambda_2}\cos\sqrt{\epsilon_r}kd_{\mathcal{I}} - \sqrt{\epsilon_r}k\lambda_1 \sin\sqrt{\epsilon_r}kd_{\mathcal{I}} \Bigr) \cosh\frac{d_{\mathcal{S}}}{\lambda_1} \nonumber \\
&&+ \Bigl( \cos\sqrt{\epsilon_r}kd_{\mathcal{I}} + \frac{\sin\sqrt{\epsilon_r}kd_{\mathcal{I}}}{\sqrt{\epsilon_r}k\lambda_2} \Bigr) \sinh\frac{d_{\mathcal{S}}}{\lambda_1} \,. 
\label{eq:denominator}
\end{eqnarray}
It should be noted that these equations are reduced to the well known expression for the semi-infinite superconductor given by $B=B_0 e^{-x/\lambda_1}$  
when the ${\mathcal{S}}$ layer and the bulk superconductor are the same material ($\lambda_1 = \lambda_2$) and the ${\mathcal{I}}$ layer vanishes ($d_{\mathcal{I}} \to 0$).

Assuming an insulator thickness $d_{\mathcal{I}} \ll 10^{-2}\,{\rm m}$, the above equations can be simplified. 
Completed results and discussions are seen in Ref.~\cite{KIS}.

\section{Summary}

The formulae that describe the RF field attenuation 
in the multilayer coating model with a single superconductor layer and a single insulator layer deposited on a bulk superconductor
were derived from a rigorous calculation with the Maxwell equations and the London equations. 
Completed results and discussions are seen in Ref.~\cite{KIS}.

\end{document}